\documentclass{aa}  
\usepackage{natbib}
\bibpunct{(}{)}{;}{a}{}{,} % to follow the A&A style
\usepackage{graphicx}
\usepackage{txfonts}
\begin{document} 
   \title{Lithium in NGC~2243 and NGC~104\thanks{Based on observations collected at the European Southern Observatory under ESO programmes 165.L-0263, 167.D-0173, 072.D-0777, 073.D-0211, 073.D-0760, 081.D-0287, 084.B-0810, 086.B-0237, 087.D-0230, 089.D-0579, 093.D-0818, 104.D-0951}}

   %%\subtitle{I. Overviewing the $\kappa$-mechanism}

   \author{M. Aoki\inst{1}\fnmsep\inst{2},
          F. Primas\inst{1},
          L. Pasquini\inst{1},
          A. Weiss\inst{3},
%          C. Charbonnel\inst{4}\fnmsep\inst{5},
          M. Salaris\inst{4},
          D. Carollo\inst{5}
          }

   \institute{European Southern Observatory, Karl-Schwarzschild-Str. 2, 85748 Garching bei M{\"u}nchen, Germany \\
              \email{maoki@eso.org}
         \and
               Ludwig-Maximilians-Universit{\"a}t M{\"u}nchen, Germany
         \and
               Max-Planck-Institut f{\"u}r Astrophysik, Karl-Schwarzschild-Str. 1, D-85748 Garching, Germany
%               \and
 %              Department of Astronomy, University of Geneva, Chemin de Pegase 51, 1290 Versoix, Switzerland
   %            \and
%IRAP, CNRS \& Univ. of Toulouse, 14, av.E.Belin, 31400 Toulouse, France
       \and
       % Institut UTINAM, CNRS UMR 6213, Univ. Bourgogne Franche-Comt\'{e}, OSU THETA Franche-Comt\'{e}-Bourgogne,
%Observatoire de Besan\c{c}on, BP 1615, 25010 Besan\c{c}on Cedex, France
%\and
Astrophysics Research Institute, Liverpool John Moores University, 146 Brownlow Hill, Liverpool L3 5RF, UK
\and
INAF - Osservatorio Astronomico di Trieste, I-34143 Trieste, Italy
             }

   \date{Received Junes, 2021; accepted 5 May, 2021}

% \abstract{}{}{}{}{} 
% 5 {} token are mandatory
 
  \abstract
  % context heading (optional)
  % {} leave it empty if necessary  
   {}
  % aims heading (mandatory)
   {Our aim was to determine the initial Li content of two clusters of similar metallicity but very different ages, the old open cluster NGC~2243 and the metal-rich globular cluster NGC~104.}
  % methods heading (mandatory)
   {We compared the lithium abundances derived for a large sample of stars (from the turn-off to the red giant branch) in each cluster. For NGC~2243 the Li abundances are   from the catalogues released by the $Gaia-$ESO Public Spectroscopic Survey, while for NGC~104 we measured the Li abundance using FLAMES/GIRAFFE spectra, which include   archival data and new observations. We took the initial Li of NGC\,2243 to be the lithium measured in stars on the hot side of the Li dip. We used the difference between the initial abundances and the post first dredge-up Li values of NGC~2243, and by adding this amount to the post first dredge-up stars of NGC~104 we were able to infer the initial Li of this cluster. 
Moreover, we compared our observational results to the predictions of theoretical stellar models for the difference between the initial Li abundance and that after the first dredge-up.}
  % results heading (mandatory)
 {The initial lithium content of NGC~2243 was found to be  A(Li)$_i$ = 2.85 $\pm$ 0.09\,dex by taking the average Li abundance measured from the five hottest stars with the highest lithium abundance. This value is 1.69\,dex higher than the lithium abundance derived in post first dredge-up stars.
   %that are found on the blue side of the Li-dip. 
%blue and the red side of the ``Li-dip'' (five on each side). 
%The difference between this value and the lithium abundance at the basis of the red giant branch ($\Delta$(Li)) of NGC~2243 is . 
By adding this number to the lithium abundance derived in the post first dredge-up stars in NGC~104, we infer a lower limit of its initial lithium content of A(Li)$_i$ = {\bf 2.34} $\pm$ 0.13\,dex. Stellar models predict similar values.
Therefore, our result offers important insights for further theoretical developments. }

  % conclusions heading (optional), leave it empty if necessary 
   {}


   \maketitle
%
%-------------------------------------------------------------------

\section{Introduction}

The evolution of lithium (Li) remains an issue not fully understood because of the incomplete knowledge of its production channels and destruction mechanisms in stellar interiors. Lithium has an important role in astrophysics, being one of the four abundance indicators (together with deuterium, $^3$He, and $^4$He) that can be used to test predictions of the Big Bang Nucleosynthesis. This is one of the main reasons why Li and its abundance in stars have been studied extensively in different types of stars and environments.

A well-known property of old metal-poor unevolved stars is their constant Li abundance \citep[known as the Spite plateau;][]{spites82, sbordone10}, which is observed in most dwarf and early subgiant stars of the Galactic halo. The typical range of stellar parameters of these plateau stars is 5800\,K$<$ $T_{\rm eff}$ $<$6800\,K and [Fe/H]$<-2$. After its first discovery many studies have confirmed the plateau \citep[e.g.][]{cp05, bonifacio07} and converged towards a Li plateau abundance A(Li)$\sim$2.2\,dex. For many years this constant value has been interpreted as the primordial Li content of the universe. 

However, the primordial level of Li is significantly higher \citep[A(Li)$\sim$2.75 dex;][]{cv17} when the primordial baryon-to-photon ratio derived from the acoustic oscillation of the cosmic microwave background measured by the WMAP satellite \citep[e.g.][]{cyburt08, larson11} and the Planck satellite \citep[e.g.][]{adam2016planck} is combined with the standard big bang nucleosynthesis (SBBN) framework. This discrepancy %or simply this difference 
has yet to be understood and solved as no major flaw or large uncertainty has been identified in the SBBN paradigm or in the observations. Even if one takes into account how easily Li can be destroyed in stellar interiors \citep[e.g.][]{tc09} due to its relatively low burning temperature ($\sim2.5 \times 10^6$ K), it is challenging to identify a depletion mechanism that would lower the WMAP$+$SBBN primordial Li abundance to a constant value of A(Li)$\simeq$2.2 over such wide range of stellar characteristics \citep[e.g.][]{richard05, richard12}. Furthermore, during the last decade it has been found that extremely metal-poor thus older halo stars ([Fe/H]$<-2.8$) tend to depart from the Spite plateau by showing even lower Li abundances \citep[e.g.][]{melendez2010, hansen2014}. 
The cause for this Li meltdown at the lowest metallicities is still under debate, and these findings indicate that the Li evolution in the early universe is not so simple, and suggest the presence of stellar nuclear processes affecting the initial Li abundance.

Cluster stars are the ideal laboratory to investigate the behaviour of a variety of chemical indicators (Li included) at different stellar evolutionary stage;  at first approximation they share the same origin and evolution, differently from halo field stars. Moreover, from accurate photometry and high-quality spectroscopic data we can infer the age and metallicity of cluster stars. 

For instance, \cite{lind09} constructed the first accurate A(Li) versus M$_{V,0}$ diagram for the metal-poor globular cluster NGC~6397 ([Fe/H] = $-$2.02) by investigating the entire evolutionary sequence, from the main sequence (MS) to the red giant branch (RGB). They found that dwarf and early subgiant branch (SGB) stars form a thin Li plateau similar to the Spite plateau, while more evolved stars follow a well-defined depletion curve characterised by some of the well-known phases in the evolution of stars \citep[e.g. first dredge-up on the SGB and additional mixing;][]{charbonnel2007thermohaline}.

NGC~104 (also known as 47\,Tuc) is a metal-rich globular cluster \citep[{[Fe/H]} = $-0.78$; e.g.][]{forbes2010accreted}. Unlike NGC~6397, NGC~104 has not been studied as thoroughly in terms of evolutionary coverage, although lithium in NGC\,104 has been studied by several authors in the past three decades \citep{pasquini-molaro1997}. \cite{bonifacio07} observed a small sample of TO stars and found them to be characterised by different Li abundances within the cluster, ranging between A(Li)= 1.58--2.14\,dex. With a significantly larger sample of turn-off (TO) stars, \cite{dorazi10} confirmed this finding and suggested that the Li content in this cluster may have been depleted in a non-homogeneous way. An independent re-analysis of this  dataset by \cite{dobrovolskas14} further confirmed star-to-star scatter in the Li abundance among TO and early SGB stars, which is larger than in similar samples in the field or in other globular clusters. They suggested that stars more metal-rich than [Fe/H]$\sim-1.0$ may experience a more significant Li depletion during their early evolutionary phases, but without identifying a specific mechanism responsible for this extra depletion. Studies of other globular cluster (e.g. M4, [Fe/H]$\sim-$1.16) also found a spread, albeit smaller, in Li abundance values among TO and early SGB stars \citep[e.g.][]{mucciarelli2011lithium}. Some authors have tentatively related the Li spread to the presence of multiple populations in NGC\,104, though no strong anti-correlation between Li and Na abundance has been found \citep{bonifacio07, dobrovolskas14}.

Similarly, studies of lithium abundances in open clusters also offer important clues to the evolution of lithium. \cite{francois13} analysed a large sample of stars in the most metal-poor open cluster, NGC~2243. They measured the metallicity to be [Fe/H]$\sim-0.54 \pm 0.10$ dex. Although Li had already been detected in four dwarf stars of this cluster by \cite{hp00}, and was found to be A(Li) = 2.35\,dex, \cite{francois13} were the first  to confirm the presence of the so-called Li dip in NGC\,2243, a marked decrease in the cluster Li content in a narrow stellar temperature range.
% (6700K--6900K). 
Since it was first observed in the Hyades open cluster \citep{boesgaard1986lithium} the Li dip feature has also been found in several open clusters, and different explanations have been proposed, from the deepening of the convection zone during the proto-main-sequence (PMS) phase of low-mass stars (M$<$1.3M$_{\odot}$) \citep[e.g.][]{cummings2017wiyn} to gravity waves \citep{tc09} that interact with rotation and influence the mixing and transport of Li in stars. 

For the purpose of this analysis, one notable aspect of NGC~2243 is its metallicity, which can be considered comparable, within the associated uncertainties, to that of NGC\,104 \citep[{[Fe/H]}$\sim-0.7$ dex; e.g.][]{carretta2009anticorrelationVIII, harris96}, although it is expected to be significantly younger. Therefore, the comparison of the Li evolutionary trends between one of the most metal-poor and oldest open clusters and a metal-rich globular cluster can shed new light on our understanding of the initial Li content as a function of age. The open cluster dwarf stars are observed with fairly high Li abundances \citep[e.g. A(Li)$\sim$2.85;][]{francois13};  those on the blue side of the Li dip are not expected to have   experienced any Li depletion yet \citep[e.g.][]{charbonnel1999hot}. Instead, the Li abundance of turn-off stars in NGC~104 have been found to be scattered, and the highest values detected are $\sim$0.3~dex lower than for TO stars in NGC~2243. A detailed comparison of the Li content of these two clusters may therefore offer new constraints on the evolution of Li and its depletion rate with time. 

%----------------------------------------------------------------- 
   \begin{figure*}
   \centering
\includegraphics[width=8.5cm]{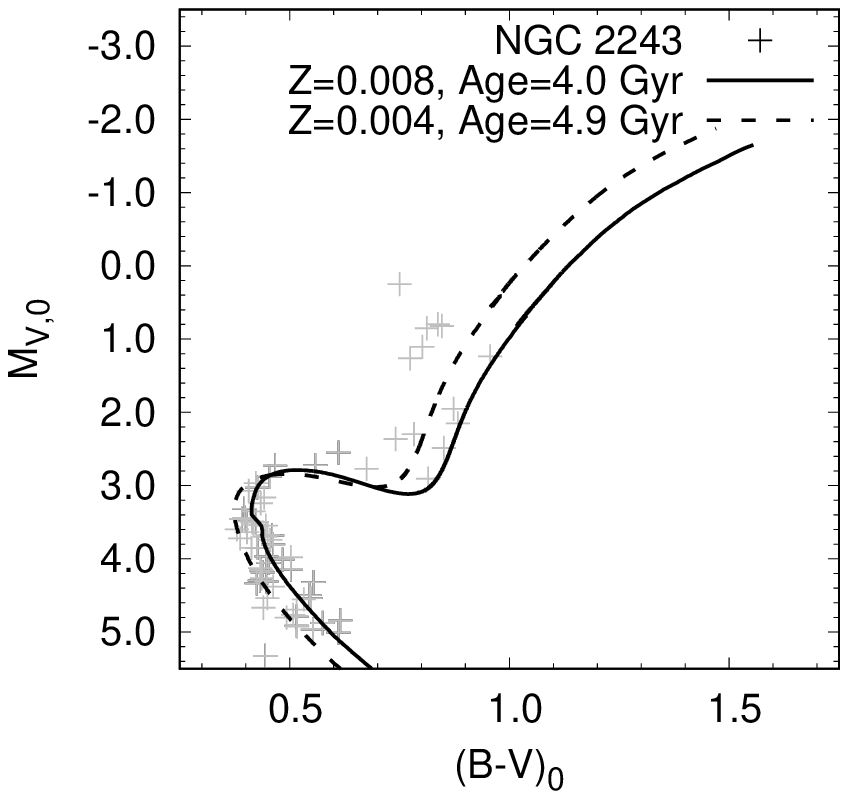}
\includegraphics[width=8.5cm]{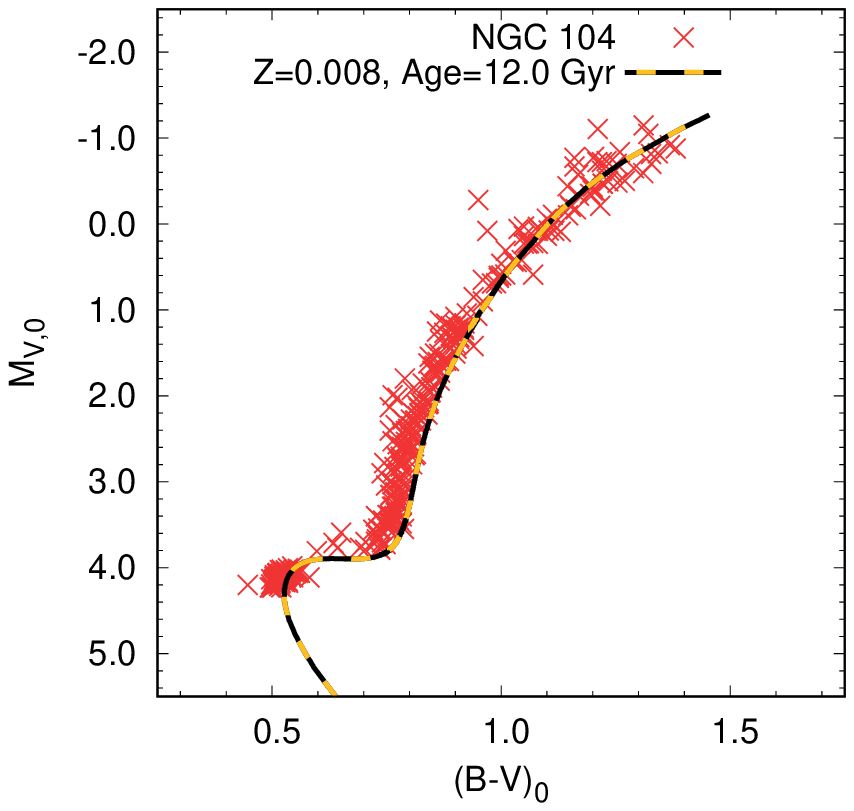}\\
\includegraphics[width=8.5cm]{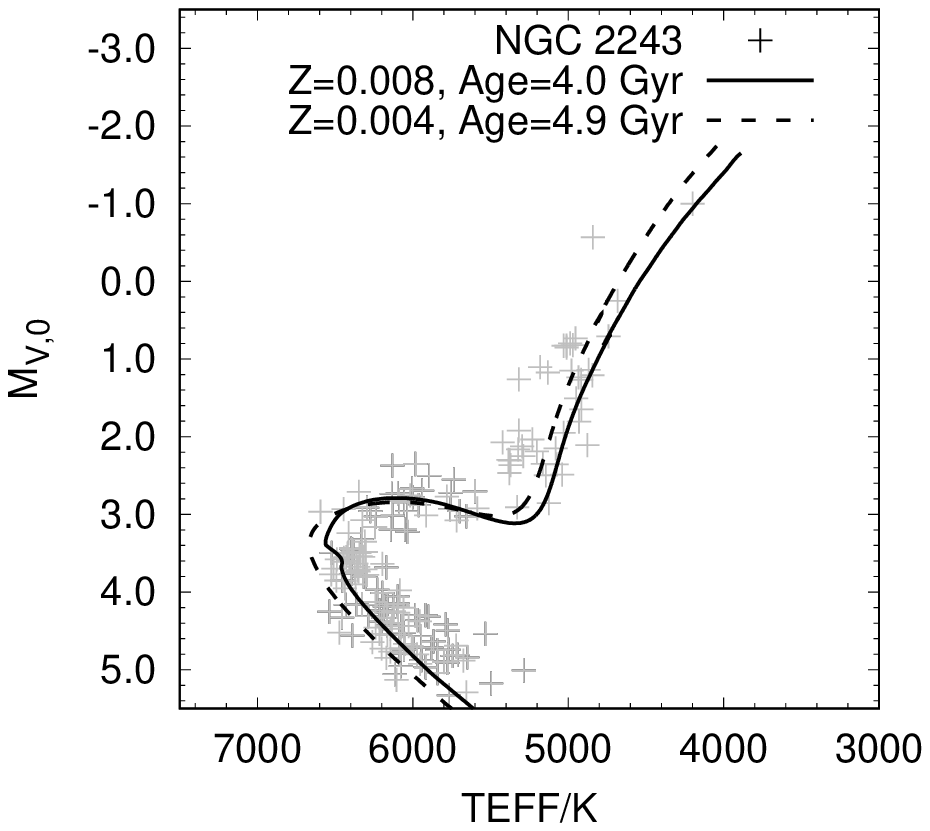}
\includegraphics[width=8.5cm]{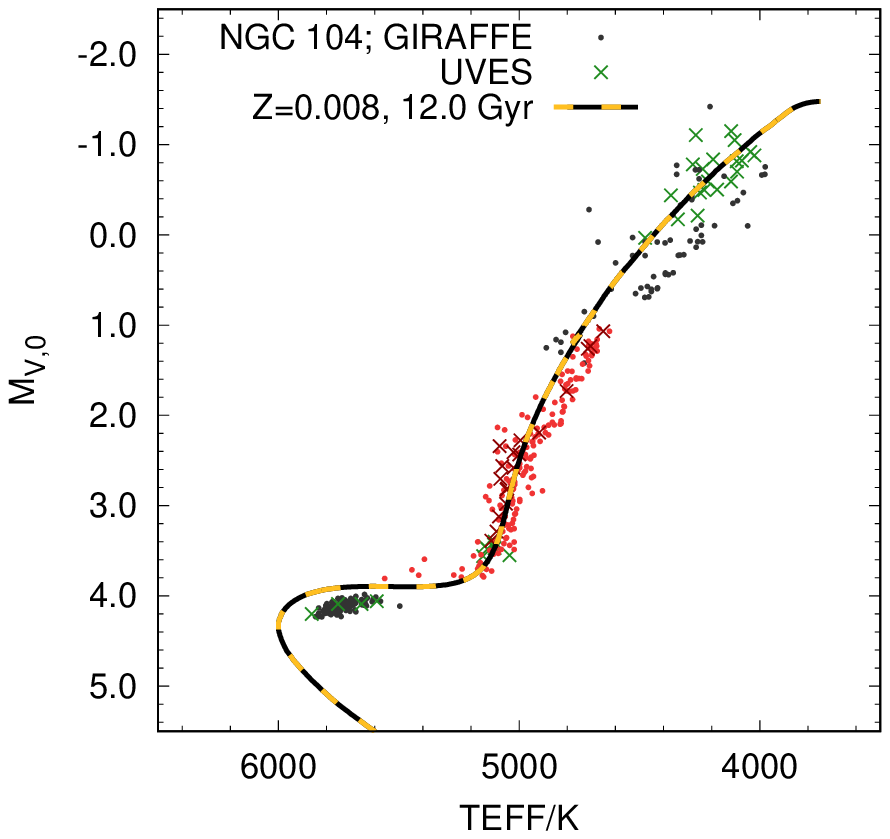}
      \caption{Colour magnitude diagrams (CMD) of NGC~2243 and NGC~104. {\it Top}:  CMD showing de-reddened (B--V)$_0$ vs. M$_{V,0}$ of NGC~2243 (left) and NGC~104 (right). {\it Bottom}:  CMD showing the adopted $T_{\rm eff}$ vs. M$_{V,0}
      $ of NGC~2243 (left) and NGC~104 (right). The lines are isochrone tracks of the clusters taken from  BaSTI. The isochrone for NGC~2243 (solid line) is computed assuming Z = 0.008, [$\alpha$/Fe] = 0.0, Age = 4.0 Gyr (corresponding to [Fe/H]=$-$0.38). Also shown (dotted line) is the track   computed assuming Z = 0.004, [$\alpha$/Fe] = 0.0, Age = 4.9 Gyr (corresponding to [Fe/H]=$-$0.69). The isochrone for NGC~104 is computed assuming Z = 0.008, [$\alpha$/Fe] = +0.4, Age = 12.0 Gyr (corresponding to [Fe/H]=$-0.68$). There are fewer stars  in the (B--V)$_0$ vs. M$_{V,0}$ diagram of NGC~2243 than its $T_{\rm eff}$ vs. M$_{V,0}$ because $Gaia-$ESO does not provide all the $B$ and/or $V$ magnitudes. The dots in the  {\it bottom right} plot show the stars observed with GIRAFFE, while the crosses  show those observed with UVES. The red symbols show our observation programme 104.D-0951(A), while the grey/green symbols show those obtained from the archive.}
               \label{fig:N2243-TeffvsMv}
   \end{figure*}
%-----------------------------------------------------------------
\section{Our target clusters: Age, metallicity, and data samples}
\subsection{NGC~2243}
NGC~2243 is a metal-poor open cluster. The first age determination was carried out by \cite{bonifazi1990ccd} using CCD photometry and found to be 4 $\pm$ 1\,Gyr. These authors also derived the metallicity of the cluster ([Fe/H] = $-$0.80 $\pm$ 0.10\,dex) and calculated its distance modulus ((m$-$M)$_0$ = 12.8 $\pm$ 0.2). Several other studies followed this work. \cite{bergbusch1991bv} derived a distance modulus of 13.05 and a cluster metallicity and age of [Fe/H] = $-$0.47 dex and 5 Gyr, respectively, using $BV$ photometry. \cite{gratton1994elemental} later confirmed this metallicity by using high-resolution spectra of two giant stars, and derived [Fe/H] = $-$0.48 $\pm$ 0.15\,dex. From $VI$ photometry \cite{kaluzny06} derived a distance modulus (m$-$M)$_V$ = 13.24 $\pm$ 0.08. The study by \cite{magrini2018gaia} derived the metallicity of the cluster [Fe/H]=$-0.38 \pm 0.04$, based on the fifth $Gaia-$ESO internal data release (GESiDR5). The recent chemical abundance studies by \cite{kovalev2019non} on the other hand derived [Fe/H]=$-0.57 \pm 0.01$.

\subsection{The $Gaia-$ESO sample}
Our NGC~2243 data sample is entirely based on the $Gaia-$ESO Public Spectroscopic Survey catalogue \footnote{https://www.eso.org/qi/catalogQuery/index/121} (hereafter GES), which collects the spectroscopic data taken up to Data Release 3 \citep[DR3;][]{gilmore2012gaia}. The GES is a large public spectroscopic survey that has systematically covered all major components of the Milky Way. The survey, now completed, aimed to provide the first homogeneous overview of the kinematical and chemical signatures in our Galaxy, using the VLT multi-object spectrograph FLAMES \citep{pasquini2003installation}. 
%The catalogue also contains observations of the most studied clusters, including NGC~2243. 
From the GES catalogue we took the NGC~2243 stars in which lithium has been detected or reported as an upper limit.  We also took the corresponding effective temperatures $T_{\rm eff}$ and $V$ magnitudes. The left  panels of Figure \ref{fig:N2243-TeffvsMv} show the colour--magnitude diagrams (CMDs) of the de-reddened (B--V)$_0$ versus M$_{V,0}$ (top) and $T_{\rm eff}$ versus M$_{V,0}$ of NGC~2243 (bottom). We employed the values of reddening ($E(B-V)=0.07\pm0.01$) and true distance modulus ($(m-M)_0=12.8\pm0.2$) provided by \cite{bonifazi1990ccd}.

The isochrones overplotted on the observed data were taken from the  Bag of Stellar Tracks and Isochrones (BaSTI) grid \citep[][]{pietri04, pietri06, pietri13}\footnote{http://albione.oa-teramo.inaf.it} and were computed for two values of the total metallicity Z of the cluster in order to cover the range of metallicities found in the literature \citep[from {[Fe/H]}=$-$0.38\,dex by][to {[Fe/H]}=$-0.80$ by \citeauthor{bonifazi1990ccd} \citeyear{bonifazi1990ccd}]{magrini2018gaia}: the isochrone computed for Z$=$ 0.008, [$\alpha$/Fe] $=$ 0.0, Age $=$ 4.0 Gyr corresponds to [Fe/H]=$-$0.35 (solid line), whereas the isochrone computed for Z $=$ 0.004, [$\alpha$/Fe] $=$ 0.0, Age $=$ 4.9 Gyr corresponds to [Fe/H]=$-$0.69 (dotted line).  Both use the solar mixture of \cite{grevesse1993atomic}.
The main purpose of this comparison was to constrain the parameters that will be used later on for the stellar models. Moreover, we chose the older version of the BaSTI grid that included alpha-enhancement because the most recent release\footnote{http://basti-iac.oa-abruzzo.inaf.it/} only provides solar scaled [$\alpha$/Fe], and we wanted to use the same grid for both clusters. 

\subsection{NGC~104}
NGC~104 is one of the most metal-rich globular clusters among the classical halo clusters, and it has been targeted by several studies. For example, one of the first studies to derive the metallicity of the cluster is by \cite{ft60}, who determined a cluster metallicity of [Fe/H]$> -0.6$ dex from the spectra of giant stars. From isochrone fitting and assuming a metallicity of [Fe/H] = $-0.48$, \cite{hvb73} derived a cluster age of 9.3 Gyr and noted that their derived metallicity was consistent with the measurement by \cite{eggen72} based on $UBV$ observations. 
\cite{sw98} determined the age of NGC~104 with $\alpha$-enhanced models and found 9.2 $\pm$ 1.0\,Gyr. \cite{gratton2003distances} derived the age of the cluster from the luminosity of the TO stars  \citep[e.g.][]{renzini1988tests}. They first derived a metallicity of [Fe/H] = $-$0.66 $\pm$ 0.04 and a distance modulus of $(m-M)_{V}$ = 13.47 $\pm$ 0.03, and then inferred an age of 10.8 $\pm$ 1.1 Gyr. 
 \cite{zoccali2001white} presented a determination of the distance (and age) of NGC\,104 by comparing the white dwarf cooling sequence of the cluster with the empirical fiducial sequence of local white dwarfs of known trigonometric parallax. They estimated the true distance modulus of $(m-M)_{0}$ = 13.09 $\pm$ 0.14, adopting the reddening $E(B-V)=0.055 \pm 0.02$. They derived the age of the cluster  to be 13 $\pm$ 2.5 Gyr.
Among the most recent studies, \cite{thompson2020erratum} measured the cluster distance and age using photometric and spectroscopic observations of the eclipsing binary in NGC~104. They derived a distance of $\sim$4.5~kpc and an age of 12.0 $\pm$ 0.5 Gyr. Chemical abundance studies \citep[e.g.][]{cordero2013detailed, kovalev2019non} have explored different elements, and they all agree that the alpha-elements are enhanced \citep[e.g. {[$\alpha$/Fe]}$\sim$0.4; ][]{kovalev2019non}.

\subsection{Archival data}
We searched the ESO Science Archive\footnote{http://archive.eso.org/wdb/wdb/adp/phase3$\_$spectral/form?} and harvested a large number of FLAMES-GIRAFFE and UVES datasets belonging to different programmes,
all targeting NGC~104. We then downloaded the most relevant (i.e. covering the spectral range of the Li line at 670.7~nm) and publicly available spectra for our analysis. In total, we analysed more than 1000 spectra, belonging to 360 individual stars. 

It should be  noted that a significant fraction of the NGC\,104 sample belongs to our own P104 observing programme (PI: Aoki). Although it was secured for a different project (Aoki et al. 2021b, in prep.), it offered us a better coverage of the Li evolutionary history of this cluster, from the TO to the upper RGB.

As it will be reported in more detail in Aoki et al. (2021b, in prep.), our PI data were reduced using EsoReflex \citep{freudling13}, which includes standard tasks such as bias subtraction, flat-fielding, and wavelength calibration. Individual exposures were then corrected for the heliocentric radial velocity by using the IRAF\footnote{IRAF is distributed by National Optical Astronomy Observatories, which are operated by the Association of Universities for Research in Astronomy, Inc., with the cooperation of the National Science Foundation.} tasks $rvid$ and $dopcor$. The final radial velocity per star was derived from the average of the individual measurements.

The mean radial velocity obtained for the cluster members is $-16.20\pm7.36$~kms$^{-1}$, where its uncertainty represents the standard deviation of the stars that contribute to the calculation of the mean. This is in good agreement with $-17.3\pm9.9$~kms$^{-1}$ \citep{wang2017sodium} and $-17.6\pm7.2$~kms$^{-1}$ \citep{dobrovolskas14}. We found only one star for which the final radial velocity was more than 3$\sigma$ away from the mean value of the cluster, and thus we discarded it from any further analysis as a likely a non-member. We also note that the radial velocities derived from GIRAFFE or UVES spectra, when both were available for the same star, are in good agreement (within $\sim 3\,$kms$^{-1}$). 

For all other (older) archival spectra, data products are readily available from the ESO Science Archive. All spectra were then normalised using the IRAF task $continuum$. At the Li line (670.7\,nm), we achieved typical signal-to-noise ratios (S/N) ranging between 40 and 80 for dwarf stars, and up to 300 for the brightest RGB stars. 

The right  panels of Figure \ref{fig:N2243-TeffvsMv} present the CMD of NGC~104, showing (B--V)$_0$ versus M$_{V,0}$ (top) and $T_{\rm eff}$ versus M$_{V,0}$ (bottom), with overplotted isochrones. The bottom right figure includes information on the instrumental source of the data (GIRAFFE versus UVES) and highlights our PI dataset.
We employed the values of reddening and distance modulus  estimated by \cite{zoccali2001white}.
We used the same release of BaSTI isochrones, and selected the isochrone computed with Z\,$=$\,0.008, [$\alpha$/Fe]\,$=$\,0.4, Age\,$=$\,12\,Gyr. We  assumed a standard model with no overshooting, diffusion, or mass loss. 

\section{Stellar parameters and Li abundances}
\label{teff-comparison-paper}
\subsection{Stellar parameters}
For NGC~104, we estimated the $T_{\rm eff}$ from $(B-V)_{0}$ photometry \citep{rami05} because this is the only photometric index available for all our stars. We employed the metallicity [Fe/H]$=-0.72\pm0.05$ taken from the globular cluster catalogue by \cite{harris96}.
For NGC~2243, the GES catalogue provided the stellar parameters and Li abundance of individual stars.

\begin{figure}
   \centering
\includegraphics[bb=0 0 510 874, width=5.0cm, angle=270]{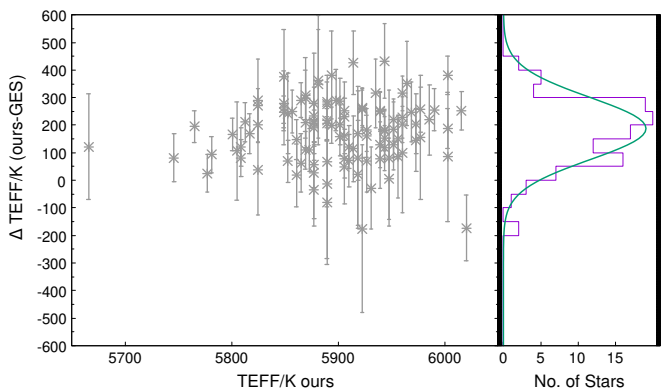}
      \caption{$T_{\rm eff}$ vs. delta-$T_{\rm eff}$ ($T_{\rm eff}-$$T_{\rm eff GES}$) of NGC~104 TO stars. Our $T_{\rm eff}$ is derived from $B-V$ photometry. The histogram on the right shows the distribution of delta$T_{\rm eff}$. The Gaussian fitting shows the maximum level at $\sim$175~K with $\sim$113~K as the $\sigma$ error.}
      \label{fig:deltaTeff}
   \end{figure}

However, since this work relies on a detailed comparison between the Li evolutionary curves of these two clusters and Li abundances are very sensitive to the stellar temperature, we carefully checked how the different $T_{\rm eff}$ scales used for the two clusters compare to each other. We performed this test on 106 stars in NGC\,104 that also appear in the GES catalogue.
Figure \ref{fig:deltaTeff} shows our $T_{{\rm eff,}(B-V)_0}$ values as a function of the $T_{\rm eff}$ difference between our data and that of the  GES, showing that on average our $T_{\rm eff}$ values are about 175\,K higher than those derived by the GES. Therefore, to properly compare  the Li content of the two clusters, we corrected our $T_{\rm eff}$ scale downwards by $-$175\,K. A similar test was done also for the evolved stars (identified here as those with $T_{\rm eff} <$5500\,K) in common with the GES catalogue (25 in total). We found that our values are about 46\,K higher than those derived by GES, therefore, we corrected our $T_{\rm eff}$ scale  downwards by $-$46\,K for stars that have effective temperatures lower than 5500\,K. The final $T_{\rm eff}$ values adopted are those displayed in the bottom right panel of Figure \ref{fig:N2243-TeffvsMv}. 
The colour calibration error on $T_{\rm eff}$ for individual stars is calculated considering the propagation of uncertainties due to reddening and metallicity. We also considered the internal scatter of the colour--temperature calibration \citep[88 and 51 K for dwarf and giant stars, respectively;][]{rami05}. The final uncertainty on the temperature is calculated from the quadrature sum of the colour calibration error and the internal scatter. The derived uncertainties are in the range  59--118K.

The surface gravity (log {\it g}) was determined by using the standard relation that combines stellar mass, $T_{\rm eff}$, and bolometric magnitude:

    \begin{equation}
\log{\it g}_{\mathrm{*}}=\log{\it g}_{\odot}+\log \frac{M_{\mathrm{*}}}{M_\odot}+4\log \frac{T_{\mathrm{eff *}}}{T_{\mathrm{eff}\odot}}+0.4 \left( M_{\mathrm{Bol*}}-M_{\mathrm{Bol\odot}}\right)
   .\end{equation}

\noindent We adopted log {\it g}$_{\odot}$ = 4.44,  $T_{\rm eff \odot}$ = 5770\,K, as $M_{\rm Bol \odot}$ = 4.74 as  representative values for the Sun. We adopted the mass of the stars as constrained by the isochrone comparison. For NGC~104, we adopted $M_{\rm *}$ = 0.90~M${_\odot}$ as the representative mass of RGB stars, $M_{\rm *}$ = 0.89~M${_\odot}$ for SGB stars, and $M_{\rm *}$ = 0.86~M${_\odot}$ for TO stars. 
%For NGC~2243, we adopt $M_{\rm *}=1.11{M_\odot}$ for the mass of the RGB stars, $M_{\rm *}=1.08{M_\odot}$ for the mass of the SGB stars and $M_{\rm *}=1.05{M_\odot}$ for the mass of the TO stars. 
Although our assumptions on the stellar masses may not be very accurate, we note that a 0.3 ${M_\odot}$ change in mass would only affect the calculated log{\it g} by 0.1\,dex. Using the calibration for the bolometric correction from \citet{alonso99} we then calculated the absolute bolometric magnitude ($M_{\rm Bol *}$) of the stars assuming the apparent distance modulus of 13.27$\pm0.14$ \citep{zoccali2001white}. We calculated the uncertainty on log{\it g} by adopting the errors on stellar mass (0.09$M_\odot$), which results in an average error of $\sim0.24$~dex. 
%(Table \ref{tab:list-stars}). 
When we change the log{\it g} by $\pm$0.25\,dex the variation in Li abundance is about 0.02\,dex for giant stars, and even less for dwarf stars.
%----------------------------------------------------------------- 
%-----------------------------------------------------------------

We adopted a single value for the microturbulence ($\xi$\,=\,1.5~km/s) for all stars. \cite{bonifacio07}, \cite{dorazi10}, and \cite{dobrovolskas14} all used  $\xi=1.0$ km/s for the analysis of dwarf stars in NGC 104. On the other hand, if we apply the microturbulence relation \citep[][applicable only to dwarf stars]{edvardsson1993chemical}, the typical $\xi$ derived is 1.65 km/s. 
Moreover, \cite{wang2017sodium} studied the RGB stars of NGC\,104 and reported  $\xi$ values in the  range 0.98--1.64\,km/s. Because of the methods we   followed for the determination of the other stellar parameters, we decided to apply a single value of $\xi$ = 1.5~km/s. We note that the effect of any change in the microturbulence on the derived Li abundance is negligible. A variation in $\xi$ of 1\,km/s affects the Li abundance by less than 0.01\,dex.

\subsection{Li abundances}
Finally, we computed the Li abundances. For NGC~2243, as previously mentioned, we took them directly from the GES DR3 catalogue. For NGC~104, instead, the Li abundances were derived from a local thermodynamic equilibrium (LTE) analysis of the 670.7~nm line, applying the spectral synthesis code based on the model atmosphere programme by \citet{tsuji78}.
The ATLAS model atmospheres with the revised opacity distribution function \citep[NEWODF][]{caskur03} were employed. 
The line list was assembled from the VALD database \citep{kupka99}. Non-local thermodynamic equilibrium (NLTE) corrections were initially applied to our NGC~104 sample following the prescriptions by \cite{lind2009departures}, but for the main purpose of this project (i.e. direct comparison with NGC~2243) all Li abundances shown hereafter are LTE because this is what the GES catalogue provides. We note, however, that the difference between LTE and NLTE values measured for our most critical group of stars (i.e. those at the post first dredge-up stage) is truly small and does not affect our conclusions. We also note that the Li abundances derived from GIRAFFE or UVES spectra, when both available for the same star, are in good agreement ($<$0.2\,dex). 
We excluded the six stars in our sample with effective temperatures lower than 4000\,K because their Li lines are buried in the TiO bands.
%----------------------------------------------------------------- 
   \begin{figure}
   \centering
\includegraphics[width=10cm]{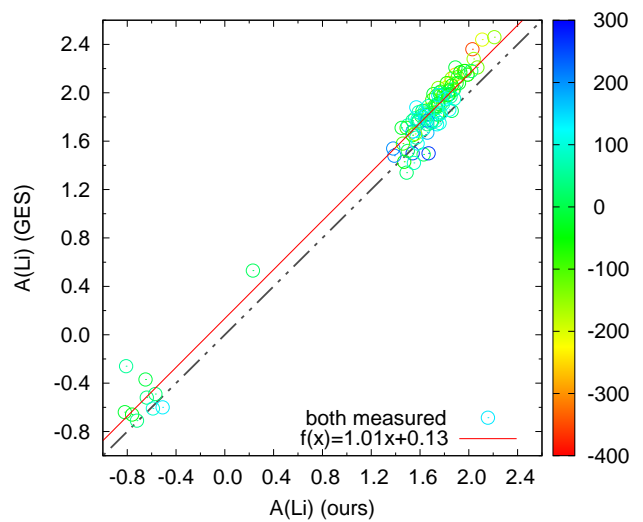}
      \caption{Comparison of A(Li)$_{\rm LTE}$ abundances in NGC\,104 between our measurements and the GES catalogue. The colour bar indicates the difference (our data $-$ GES) in effective temperature for the stars in common. The red solid line is the fit to the data points.}
      \label{fig:LivsLi-N104}
   \end{figure}
%-----------------------------------------------------------------

We then performed one further consistency test between the Li abundance scales of the two clusters by comparing our Li measurements in NGC\,104 to those reported in the GES catalogue for the same globular cluster. We note that the GES catalogue has a smaller number of NGC\,104 stars compared to our total sample, which explains why we could not use the GES abundances, as done for NGC\,2243. From Figure \ref{fig:LivsLi-N104} it is evident that despite the correction already applied to our temperature scale, there is still, on average, a 0.13 dex difference between our data and that of the  GES. This offset could be due to remaining differences in the temperature of individual stars, in the spectral synthesis code and/or the model atmospheres used in the two analyses. In order to be as close as possible to the GES abundance scale also for NGC\,104, we  decided to apply this small correction ($+$0.13\,dex) to our Li measurements in NGC\,104,  noting however that this difference is well within our estimate of the overall uncertainty of our Li measurements.

\begin{figure}
   \centering
\includegraphics[width=8.5cm]{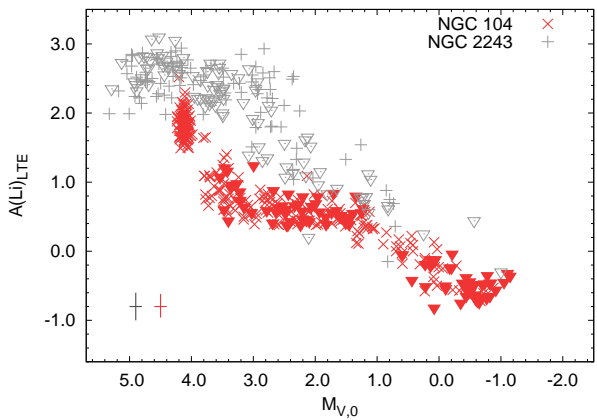}
      \caption{A(Li) as a function of M$_{V,0}$ of NGC~2243 and NGC~104. The plus signs represent the NGC~2243 sample, while crosses refer to NGC~104 (with the $+$0.13\,dex upward correction). Li upper limits are shown by open and filled upside-down triangles, grey and red, respectively. At the lower left corner, mean error bars are shown for NGC\,2243 (in grey) and NGC\,104 (in red). }
               \label{fig:N2243-LivsMv}
   \end{figure}

Figure \ref{fig:N2243-LivsMv} shows A(Li) as a function of M$_{V,0}$, where the red symbols represent our corrected ($+$0.13\,dex) Li abundances in NGC~104, while the grey symbols represent the GES abundances in NGC~2243.
%{\color{blue}[ADD 0.13 dex]}}

We note that this study is the first to show the evolution of the Li abundance in NGC~104 from the MS-TO to the upper RGB. We confirm the presence of a Li spread among TO stars, as already observed by \cite{dorazi10} and \cite{dobrovolskas14}, but this is only marginally larger than the  observational errors. Since it is not the aim of this work to discuss the nature of the Li spread, we refer the reader to Aoki et al. (2021b, in prep.). %{\color{red}[Aoki et al. (2021b) - pls insert ref in the right way]} 
that will describe in detail the analysis of this dataset. 

\subsection{Error estimates}
%-----------------------------------------------------------------------------------------------------------------------------
\begin{table}
\begin{center}
\caption{Li abundance changes (dex) by changing $T_{\rm eff}$}
\begin{tabular}{lrr}
\hline
\hline
            \noalign{\smallskip}
& \multicolumn{2}{c}{$\Delta$ $T_{\rm eff}$ } \\
\cline{2-3} 
            \noalign{\smallskip}
$T_{\rm eff}$ range & $+100$K & $-100K$ \\
\hline
            \noalign{\smallskip}
%~~~~~~~$<$~4000&+0.14&$-0.14$\\
4000--4500&+0.18&$-0.18$\\
4500--5000&+0.16&$-0.17$\\
5000--5500&+0.12&$-0.12$\\
5500--6000&+0.08&$-0.09$\\
%6000~$<$~~~~~~~&+0.07&$-0.07$\\
\hline 
 \end{tabular}
%\tablefoot{The difference is taken as the abundance measured after changing the $T_{\rm eff}$ minus our final abundance of Li. {\color{red}[FP: what does this footnote mean? Is it needed?]}\\}
\label{tab:error}
\end{center}
\end{table}
%-----------------------------------------------------------------------------------------------------------------------------
The Li abundances are based on one neutral atomic line; therefore, it is mostly sensitive to the choice of the stellar effective temperature, which represents the largest source of error.
We investigated the systematic difference in the Li abundances that occurs when uncertainties on the derived $T_{\rm eff}$ are taken into account. Table~\ref{tab:error} reports our findings. We grouped our sample in temperature bins and chose five sample stars from each bin. We re-derived the Li abundances for these five stars by changing their $T_{\rm eff}$ by $\pm$100\,K. The Li abundance increases with higher $T_{\rm eff}$, and the difference in Li is larger for lower temperature stars.
{Any dependence on the other stellar parameters was ignored since the typical errors given for log${\it g}$ and $\xi$ have negligible effects on the Li abundance, and the error on [Fe/H] is already included in the error on $T_{\rm eff}$.}

We also took into account the uncertainty on the continuum placement  estimated from the neighbouring portions on both sides of the Li line (more specifically, from 6706.9--6707.5\AA\ and 6708.3--6708.8\AA). 
%We tested the continuum levels with different assumptions, varying the values of the continuum levels slightly, then observe the difference in Li abundance. 
We tested our choice of the continuum placement by varying it slightly (by 1--2\%) and observing its effect on the derived Li abundance. 
%When the variation in the continuum level differs by more than the Li line depth, we define them as Li upper-limits.
When the shift in the continuum placement differs by more than the depth of the Li line, we considered the measurement of these lines as upper limits. 
The typical uncertainty ranges from 0.04 to 0.16\,dex, depending on the S/N or the resolution of the spectra. The total error on the Li abundance is calculated from the quadrature sum of the Li error estimated from $\pm100$\,K variation of $T_{\rm eff}$ (given in Table \ref{tab:error}) and the error associated with the placement of the continuum level, and these sum up to $\pm$0.20\,dex and $\pm$0.16\,dex for NGC\,2243 and NGC\,104, respectively.

Table \ref{tab:list-stars}\footnote{Full table available electronically at the CDS.} lists the stars of our NGC\,104 sample that are confirmed cluster members together with their coordinates, photometry, radial velocities, S/N, final effective temperatures, log{\it g} values, and Li abundances.
%{\color{blue} [ADD 0.13 dex]}

%-----Stellar Parameter Table-----------------
\begin{table*}
\caption{\label{tab:list-stars} NGC~104 cluster members together with their coordinates, photometry, radial velocities, S/N, effective temperatures, log{\it g} values, and measured Li abundances. The full table can be retrieved from the CDS.}
\centering
\begin{tabular}{lccccccccc}
\hline\hline
            \noalign{\smallskip}
ID & RA & DEC & V & B-V & Teff & log{\it g} & S/N & RV  & A(Li)$_{\rm LTE}$ \\
 & (hh:mm:ss.ss)  & (dd:mm:ss.s) &  &  & [K]  &[dex] &(6708\AA)  &[kms$^{-1}$] &[dex]\\
\hline
            \noalign{\smallskip}
429&00:20:18.60 &-72:01:11.40 &16.70&0.80&5104 $\pm$ 83 &3.55 $\pm$ 0.24 &48.30&-19.57 $\pm$ 0.33 &1.40 $\pm$ 0.14  \\
433&00:21:49.12 &-72:02:52.60 &16.72&0.78&5143 $\pm$ 84 &3.57 $\pm$ 0.24 &55.20&-22.87 $\pm$ 0.42 &0.77 $\pm$ 0.14  \\
435&00:21:14.07 &-72:03:33.20 &16.74&0.80&5102 $\pm$ 83 &3.56 $\pm$ 0.24 &54.70&-20.17 $\pm$ 0.65 &1.35 $\pm$ 0.14  \\
456&00:21:30.42 &-72:05:35.90 &16.75&0.79&5131 $\pm$ 84 &3.58 $\pm$ 0.24 &52.50&-16.19 $\pm$ 0.39 &0.60 $\pm$ 0.14  \\
478&00:21:08.14 &-71:58:47.10 &16.83&0.78&5150 $\pm$ 84 &3.62 $\pm$ 0.24 &42.40&-20.01 $\pm$ 0.31 &1.09 $\pm$ 0.15  \\
...&... &...&...&...&... &... &...&... &...  \\
\hline
\end{tabular}

\end{table*}
%------------------------------------------------------------------------------------------------------

%----------------------------------------------------------------- 
%-----------------------------------------------------------------\\

\section{The initial lithium content of NGC~2243 and NGC~104}\label{mucciarellimethod}
%----------------------------------------------------------------- 
   \begin{figure*}
   \centering
         \includegraphics[width=8.5cm]{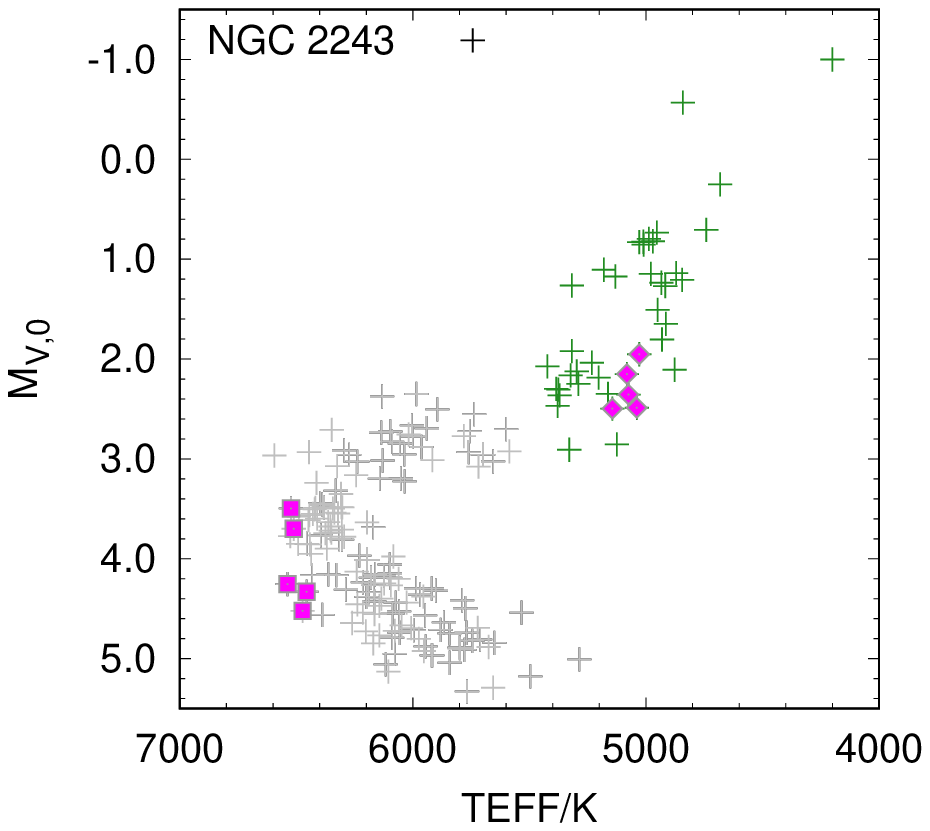}
\includegraphics[width=8.5cm]{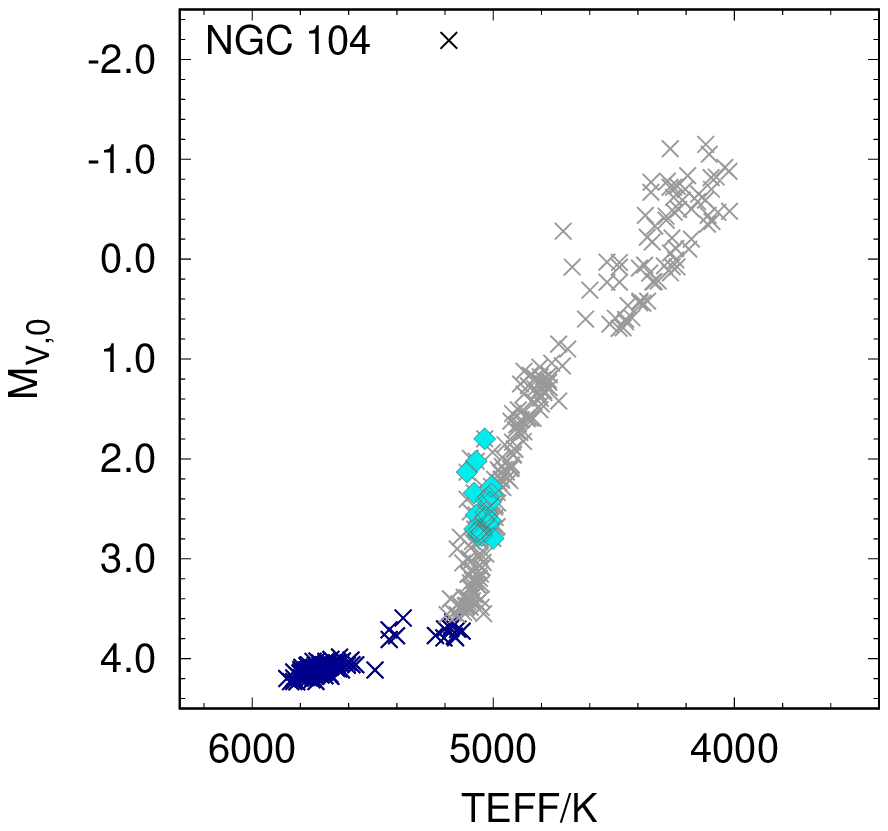}
      \includegraphics[width=8.5cm]{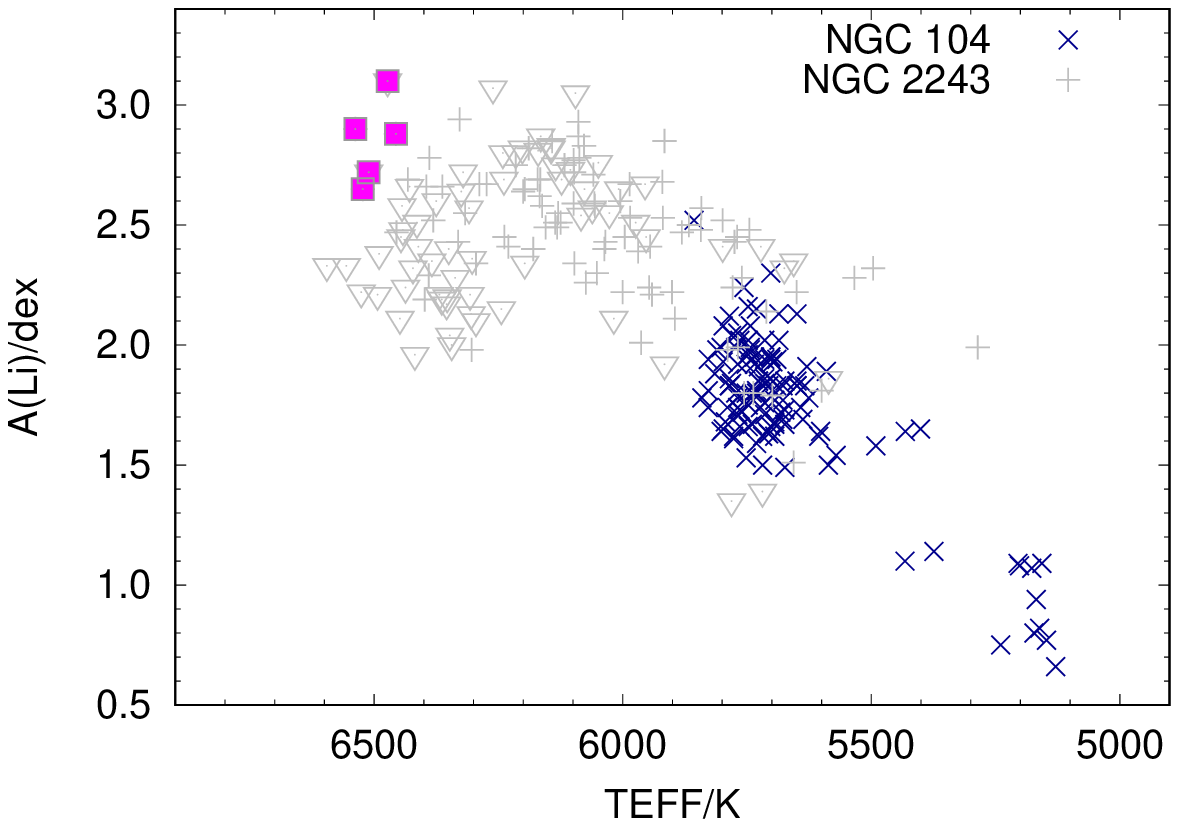}
\includegraphics[width=8.5cm]{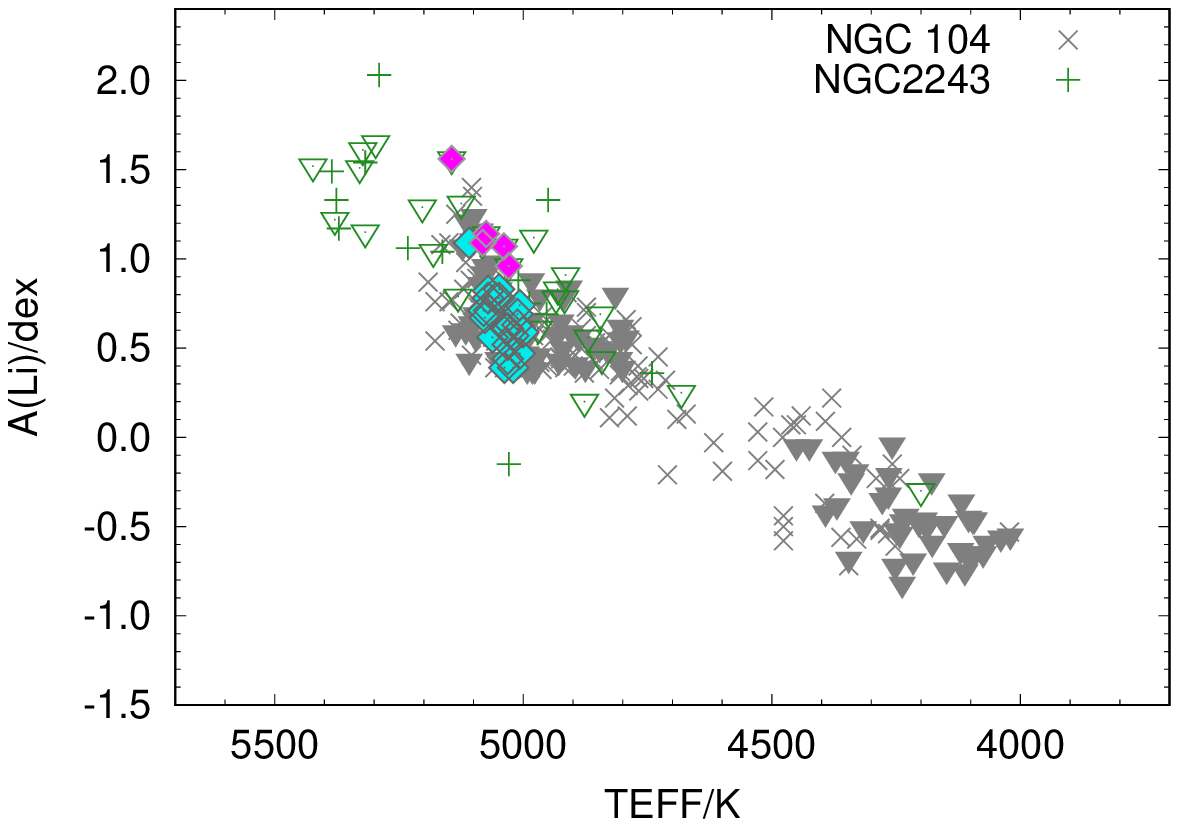}
  \caption{Stars of NGC~2243 and NGC~104 showing M$_{V,0}$ and A(Li) as a function of $T_{\rm eff}$. {\it Top:} $T_{\rm eff}$ vs. M$_{V,0}$ of the NGC~2243 ({\it left}) and NGC~104 ({\it right}). The grey symbols in the left plot show the NGC~2243 stars up to the basis of the RGB (un-evolved stars), while the green symbols show  the evolved stars after the basis of the RGB (evolved stars). The pink squares   show the five highest Li abundance stars located at the highest $T_{\rm eff}$ of NGC~2243. The pink diamonds   show five stars identified as the `basis of the RGB', and selected with constraints of  $5000< T_{\rm eff} <5200$ and $1.5<$M$_{V,0}<2.5$. The right plot shows  NGC\,104 stars. The blue symbols are  `un-evolved' stars, while the grey symbols are `evolved' stars. The light blue diamonds  show 22 stars identified as the `basis of the RGB' for NGC~104, and selected with constraints of  $5000<T_{\rm eff} <5200$ and 
  $1.8<$M$_{V,0}<2.8$.
{\it Bottom:} $T_{\rm eff}$ vs. A(Li) of NGC~2243 and NGC~104. Colours and symbols are the same as in the top two  graphs. The upside-down triangles show the upper limit of the Li abundance. 
Temperatures and Li abundances of NGC\,104 stars include the $-$175\,K and $+$0.13\,dex  corrections (see text).}
         \label{fig:LiTeff-BF}
   \end{figure*}
%-----------------------------------------------------------------
\cite{mucciarelli12} studied the evolution of surface Li abundances in post first dredge-up stars (pFDU) covering a wide range of metallicity ($-3.62<$[Fe/H]$<-1.01$). They selected 17 metal-poor lower RGB field halo stars ([Fe/H] $<-1$\,dex) fainter than the RGB bump magnitude, and derived accurate metallicities and Li abundances. 
By correcting their Li abundances for the amount of dilution predicted by stellar evolution models (computed as the difference $\Delta$(Li) between the models initial and pFDU Li surface abundances), they inferred the initial amount of lithium of these lower RGB halo field stars. 
Their study took the combined  WMAP and SBBN predicted value of A(Li)=2.72\,dex as the initial lithium of the stellar models, which they noted that this choice has no impact on the predicted $\Delta$(Li) values, and assumed that 12.5\,Gyr RGB stars have stellar masses between 0.80 $-$ 0.86 M$_{\odot}$, increasing with Z. 

They used three different sets of theoretical models (with and without diffusion, and a third  without diffusion but accounting for the overshooting below the convective envelope) and found that the amount of $\Delta$(Li) increases with increasing metallicity, reaching values between 1.44 and 1.51\,dex (depending on the model) at the higher metallicity end of their models ([Fe/H] = $-$1.01). One could thus expect that a cluster at even higher metallicity has an even larger $\Delta$(Li). The $\Delta$(Li) predicted from the models with diffusion are larger by $0.07$\,dex with respect to those derived from the models without diffusion at any metallicity. The models including overshooting predict $\Delta$(Li) larger by 0.01\,dex with respect to the model without diffusion at [Fe/H]=$-3.62$, while the difference increases to 0.06\,dex at [Fe/H]=$-1.01$. Overall, the model with diffusion predicts the largest $\Delta$(Li). 

Figure \ref{fig:LiTeff-BF} shows the A(Li) versus $T_{\rm eff}$ trends for NGC~2243 and NGC~104 (bottom graphs) but separated in the two phases we are most interested in: MS, TO, and SGB stars (i.e. all stars before the base of the RGB; bottom left) and RGB stars (i.e. those from the base of the RGB and above; bottom right). The hottest stars in NGC~104 have lower Li abundances compared to the stars in the same temperature range of NGC~2243.
We note that \cite{dorazi10} and \cite{dobrovolskas14} classified these stars as TO stars, but we are not confident about their exact evolutionary status. A closer inspection of the right panels of Figure \ref{fig:N2243-TeffvsMv} show that they are indeed TO stars in the M$_{V,0}$ versus ($B-V$)$_0$ graph, but look more like subgiants in the M$_{V,0}$ versus $T_{\rm eff}$ graph. If the latter, then these stars may have already experienced some dredge-up and their Li abundances would represent only a lower limit of the TO Li abundance. This uncertainty is  kept in mind during the  discussion of our dilution factors.

In this work, we attempt to determine the $\Delta$(Li) dilution factors empirically, by using high-resolution stellar spectra of stars in NGC~2243. 
We selected the stars with the highest $T_{\rm eff}$ (pink squares, 
%on the top right graph of Figure \ref{fig:LiTeff-BF}, 
around $T_{\rm eff}$ $\sim$ 6500\,K) as representative of the initial Li content of NGC~2243. They are located on the warm side of the Li dip, hence they are expected not to have   depleted any Li yet as they are too hot and are still on the MS, a feature already seen in other young clusters such as the Hyades \citep[e.g.][]{balachandran1995lithium}.
We selected five stars with temperatures above $T_{\rm eff}$ $=$ 6450\,K and Li abundances greater than 2.5\,dex.

The average A(Li) abundance among the five hottest stars is 2.85 $\pm\,0.09$\,dex, in agreement with A(Li)=2.70 $\pm$ 0.2\,dex measured by \cite{francois13} and with A(Li)=2.96 $\pm$ 0.06\,dex recently derived by \cite{randich2020gaia} using  GESiDR5, which is not yet publicly available.

Following \cite{mucciarelli12}, we then derived the average Li abundance in pFDU stars (Figure \ref{fig:LiTeff-BF}, top graphs), where we expect to see the effect of the first dredge-up (i.e. a dilution of the Li content due to the deepening of the convective zone). 

Given that the NGC~104 has a total metallicity Z similar to NGC~2243 we made an initial assumption that the two clusters have experienced similar amounts of dilution. By deriving the average Li abundance in pFDU stars of NGC\,104 and correcting it for the $\Delta$(Li) empirically derived in NGC\,2243, we can thus tentatively infer the initial Li content of NGC\,104. The base of the RGB in NGC~104 is easy to identify with the start of the plateau that extends from M$_{V,0}\sim$3.0 to M$_{V,0}\sim$1.4 (see Figure \ref{fig:N2243-LivsMv}). We identified 22 stars (see Figure \ref{fig:LiTeff-BF}) that fulfil the 5000$<T_{\rm eff}<$5200 and 1.8$<$ M$_{V,0}<$2.8 criteria, and for which the Li abundance could be accurately measured. The magnitude range was selected by inspecting the abundance plateau (Fig. \ref{fig:N2243-LivsMv}).
We found their average A(Li) abundance to be 0.65 $\pm$ 0.03\,dex.

For NGC\,2243 it is more complicated because the data do not show a clear plateau. We therefore decided to base our selection on the theoretical plateau (Fig. \ref{fig:N2243-LivsMv-Salaris}), which will be discussed in the following sections. This choice yielded five stars (pink diamonds in the top left panel of Fig. \ref{fig:LiTeff-BF}) that cover the ranges 5000\,K$<T_{\rm eff}<$ 5200\,K and 1.5$<$M$_{V,0}<$2.5, and for which we derived an average A(Li) value of 1.16$\pm$0.10\,dex.
The observed $\Delta$(Li) for this cluster is therefore 1.69 $\pm$ 0.13\,dex (A(Li)${_i} -$ A(Li)$_{pFDU}$), and this value is indeed larger than $\Delta$(Li) = 1.5 inferred by \cite{mucciarelli12} for their highest metallicity simulation, giving further support to their statement that the $\Delta$(Li) parameter is expected to increase with metallicity.

Finally, by applying to NGC\,104 the $\Delta$(Li) derived for NGC~2243 (i.e. 1.69\,dex), we infer an initial Li content for NGC\,104 of A(Li)$_i$ = 2.34$\pm$0.13 dex. 

\section{Observations versus theoretical predictions} 
The initial lithium tentatively inferred for NGC~104 is based on the assumption that metallicity is the most important parameter. Because  the two clusters are very similar in total metallicity, it is   reasonable to expect that their stars have undergone a similar amount of dilution until they reached the base of the RGB. On the other hand, we know that the clusters differ significantly in age and therefore also in the masses of their TO stars. Our attempt to find the best-fit BaSTI theoretical isochrones for each cluster CMD (Figure \ref{fig:N2243-TeffvsMv})   allowed us to estimate typical TO masses of 1.10~M$_{\odot}$ and 0.86~M$_{\odot}$ for NGC\,2243 and NGC\,104, respectively. Thus,  the rate at which each cluster dilutes its initial Li also may differ. Therefore, in order to validate our approach, we compared our observations to theoretical model predictions.  

Starting from the model grid of \citet{pietri04, pietri06}, we computed new stellar models using the same input physics and code as for the BaSTI isochrones displayed in Figure \ref{fig:N2243-TeffvsMv}. In the following, we   refer to these models as the `dedicated BaSTI' models, for short. 
The models include pre-MS Li depletion, but no thermohaline mixing. The latter does not affect our work, as thermohaline mixing becomes effective only after the completion of the first dredge-up (and usually accounts for the decrease after the second Li plateau around M$_{V,0}\sim$1.0). 
We chose models without diffusion since for open clusters like NGC\,2243 diffusion has a very large depletion effect on the Li abundances predicted for TO stars that is not supported by the observations. This solution requires that the gravitational settling, which is too efficient, must be carefully balanced with extra radiative levitation and rotation, but these parameters are not readily included in the model calculations. Therefore, we only apply models without diffusion for NGC\,104 and NGC\,2243.
%----------------------------------------------------------------- 
   \begin{figure}
   \centering
\includegraphics[width=9cm]{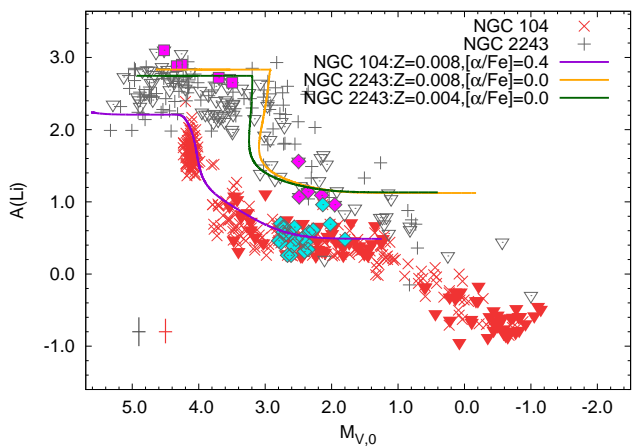}
      \caption{A(Li) vs. M$_{V,0}$ for NGC~2243 (grey plus signs and triangles) and NGC~104 (red crosses and triangles). Li upper limits are shown by upside-down triangles. Filled squares (pink) and filled diamonds (pink and light blue) are as described in  Figure \ref{fig:LiTeff-BF}. Model for NGC104: Z=0.008, M$_i$=0.85 M$_{\odot}$, [$\alpha$/Fe] = 0.4 (purple line). Models for NGC2243: Z=0.008, M$_i$=1.18 M$_{\odot}$, [$\alpha$/Fe]=0.0 (orange line) and Z=0.004, M$_i$=1.05 M$_{\odot}$, [$\alpha$/Fe]=0.0 (green line). All models are without diffusion. 
      }
               \label{fig:N2243-LivsMv-Salaris}
   \end{figure}
%-----------------------------------------------------------------
%-----------------------------------------------------------------------------------------------------------------------------
\begin{table}
\begin{center}
\caption{Predicted cluster initial Li and $\Delta$(Li) derived from the models}
$$
\begin{tabular}{lccr}

\hline
\hline
            \noalign{\smallskip}
Cluster&A(Li)$_i$&$\Delta$(Li) &Model\\
&(dex)&(dex)&\\
\hline
           \noalign{\smallskip}
%NGC\,104&{\it 2.29}&{\it1.65}&{\it Obs}\\
%NGC\,2243&2.85$\pm0.09$&1.65$\pm0.15$&Obs\\
%\hline
            \noalign{\smallskip}
%NGC\,104&1.38&2.02&Mod1\\% 2.77-0.58-0.91 2.12:2.19 1.21:1.28 2.77:2.7 1.57:1.50
%NGC\,2243&1.50&2.70&Mod2\\%2.86-0.16-1.20
%\hline
 %           \noalign{\smallskip}
%NGC\,104&1.81&2.45&Mod1\\%If PMS 2.72-0.25-0.91 0.25->0.16 1.56:1.65 2.47:2.56 2.27:2.36
%NGC\,104&2.38&1.74&Mod1\\%If PMS 2.72+0.13-1.20 0.13->0.06, 1.65:1.58 2.85:2.78  2.82:2.75(before referee)
NGC\,104&2.61&1.96&Mod1\\%If PMS 2.72+0.13-1.20 0.13->0.06, 1.65:1.58 2.85:2.78  2.82:2.75
\hline
           \noalign{\smallskip}
%NGC\,2243&2.81&1.61&Mod2\\%If PMS 2.72-0.32-0.91 0.32->0.25 1.49:1.56 2.40:2.47 2.19:2.23(before referee)
NGC\,2243&2.90&1.74&Mod2\\%If PMS 2.72-0.32-0.91 0.32->0.25 1.49:1.56 2.40:2.47 2.19:2.23
%NGC\,2243&2.75&1.55&Mod3\\%If PMS 2.72+0.03-1.20  1.62:1.55 2.82:2.75 2.69:2.71 (before referee)
NGC\,2243&2.81&1.65&Mod3\\%If PMS 2.72+0.03-1.20  1.62:1.55 2.82:2.75 2.69:2.71 (before referee)
\hline
\hline 
 \end{tabular}
 $$
 \tablefoot{ $\Delta$(Li)=A(Li)$_i$ $-$ A(Li)$_{pFDU}$, where A(Li)$_{pFDU}$ = 0.65 and 1.16\,dex, for NGC~104 and NGC~2243, respectively.
 %(ii) Values in {\it italic} (for NGC\,104) represent amounts that have been inferred based on the measured $\Delta$(Li) of NGC\,2243. The error bars represent the standard deviation among the stars that contribute to the calculation of the mean. 
  In   Col. 4 Mod 1,2, and 3 identify the different stellar models as listed in the References below. \\
}
\tablebib{
%(1)~\cite{Lagarde2017-model, Lagarde2019-model}. Z${_s}=$0.004, [$\alpha$/Fe] $=$ 0.3, M$_i$=0.9 M$_{\odot}$\\
%(2)~\cite{Lagarde2017-model, Lagarde2019-model}. Z${_s}=$0.004, [$\alpha$/Fe] $=$ 0.0, M$_i$= 1.10\,M$_{\odot}$\\
%(1)~\cite{pietri04, pietri06} with diffusion. Z=0.008, [$\alpha$/Fe] $=$ 0.4, M$_i$=0.85 M$_{\odot}$\\
(1)~\cite{pietri04, pietri06} without diffusion. Z=0.008, [$\alpha$/Fe] $=$ 0.4, M$_i$=0.85 M$_{\odot}$\\
(2)~\cite{pietri04, pietri06} without diffusion. Z=0.008, [$\alpha$/Fe] $=$ 0.0, M$_i$=1.18 M$_{\odot}$\\
(3)~\cite{pietri04, pietri06} without diffusion. Z=0.004, [$\alpha$/Fe] $=$ 0.0, M$_i$=1.05 M$_{\odot}$\\
}
\label{tab:model}
\end{center}
\end{table}
%-----------------------------------------------------------------------------------------------------------------------------

For NGC\,104 we chose a model without diffusion computed for Z=0.008, M$_i$=0.85 M$_{\odot}$ and [$\alpha$/Fe] = 0.4, which corresponds to [Fe/H]=$-0.68$ with the opacities and metal mixture of \cite{grevesse1993atomic}. 
For NGC~2243 we chose models with Z = 0.008, M$_i$ = 1.18 M$_{\odot}$,  and [$\alpha$/Fe] = 0.0 (corresponding to [Fe/H]=$-$0.38), and with Z = 0.004, M$_i$ = 1.05 M$_{\odot}$, and [$\alpha$/Fe] = 0.0 (corresponding to [Fe/H]=$-$0.69). 
Both models for NGC\,2243 include convective overshooting, as described in \cite{pietri04}, but not diffusion.

Figure\,\ref{fig:N2243-LivsMv-Salaris} shows the A(Li) versus M$_{V,0}$ diagram of the two clusters, with the dedicated BaSTI models overplotted, which for the purpose of our comparison are meaningful only starting from the TO because they have been computed for the clusters' TO masses.
The purple track is the model without diffusion computed for NGC\,104, whereas the orange and green tracks represent the models without diffusion computed for NGC\,2243 for Z=0.008 and Z=0.004, respectively.

We then computed the theoretical $\Delta$(Li) by subtracting the pFDU Li abundance from the initial Li abundance (see Table~\ref{tab:model}, Col. 2) taken from each model.
We obtained $\Delta$(Li) = 1.96\,dex for NGC\,104, and $\Delta$(Li)=1.74\,dex and 1.65\,dex for the NGC\,2243 models with Z=0.008 and Z=0.004, respectively. By comparing the models with Z=0.008 for each cluster, we found that the theoretical $\Delta$(Li) actually differs by 0.22\,dex, thus indicating that the two clusters may undergo slightly different amounts of depletion.

\section{Discussion}
The main purpose of comparing our observational Li sequences to the theoretical predictions was to test the soundness of our approach and whether we can infer the initial lithium content of NGC~104 from the $\Delta$(Li) empirically derived for NGC\,2243. For the sake of clarity, all remarks that follow (unless noted otherwise) refer to the Z=0.008/no diffusion model for both clusters. 

Within the associated uncertainties our findings are encouraging. 
Firstly, our empirically derived value of $\Delta$(Li) =  1.69$\pm$0.13\,dex for NGC\,2243 closely matches  the model prediction.
Secondly, our derived value is larger than the $\Delta$(Li) = 1.5\,dex reported by \cite{mucciarelli12} for their highest metallicity simulation (one-tenth solar). Thirdly, \cite{anthony2018wiyn} measured the Li abundance in an even higher metallicity open cluster NGC\,2506 ([Fe/H]=$-0.27$\,dex), and found average A(Li) values of 3.04\,dex (from dwarf stars, but measured on the  brighter side of the Li dip) and of 1.25\,dex at the second plateau. From these values we derive $\Delta$(Li) = 1.79\,dex. This value is still compatible with what we found for NGC2243 (1.69$\pm$0.13\,dex), although as expected it is a little higher (even with respect to the predictions of our models). We can thus state with more confidence that the $\Delta$(Li) parameter indeed seems to increase with metallicity.

For NGC\,2243, the difference between our observations-based and model-predicted values of the cluster initial lithium is  only 0.05\,dex, %0.04(before referee).
which is significantly smaller than the error associated with our %empirically derived
observation-based value ($\pm 0.09$\,dex). Since we measured the initial Li content (A(Li)$_i$ = 2.85\,dex) from MS stars on the warm side of the Li dip, this very good agreement confirms that they have depleted very little, if any, of their lithium. This is also confirmed by the negligible amount of PMS depletion predicted by the stellar models. Therefore, for NGC\,2243 we can safely conclude that we have derived a robust estimate of its initial Li content, and in turn also of the $\Delta$(Li) between its MS/PMS and pFDU stars. 

By applying our original and empirical approach to NGC\,104 instead, we found A(Li)$_{i}$ = 2.34 $\pm 0.13$\,dex, whereas the dedicated BaSTI models predict A(Li)$_{i}$ = 2.61\,dex. 
Despite this difference, we note that our A(Li)$_{i}$ inferred value
%(2.21...or 2.34)
%roughly 
matches the values predicted by the \cite{mucciarelli12} simulations (2.28$-$2.46\,dex, depending on the inclusion  or not of atomic diffusion and on the $T_{\rm eff}$ scale), who already confirmed the well-known and important discrepancy with the WMAP$-$SBBN prediction. 
We also note that our A(Li)$_{i}$ inferred value is similar to those reported by \cite{mucciarelli12} for other globular clusters like NGC\,6397, NGC\,6752, and M\,4 (i.e. A(Li)$_i\sim$2.4,  2.2, and 2.4\,dex, respectively), despite their different metal content. These findings seem to indicate that their Li content on the MS was possibly depleted by the same or a similar mechanism, more or less irrespective of metallicity. 
%@@@
However, it is also important to note that the theoretical $\Delta$(Li) values of our two clusters (see Table~\ref{tab:model}) actually differ by 0.22\,dex, which may indicate that, due to their different ages and therefore TO stellar masses, the two clusters may indeed deplete lithium in different amounts. If we were then to apply this amount as a first-order correction to account for the age difference between NGC\,2243 and NGC\,104, we would then infer an initial Li content of A(Li)$_{i}$ = 2.56 $\pm 0.13$\,dex for NGC\,104, which confirms the discrepancy with the WMAP$-$SBBN prediction, but it is in very good agreement with the A(Li)$_{i}$ predicted from the dedicated BaSTI models. We note that this can only be considered   a first-order correction because   more detailed comparisons with more detailed models are needed  to confirm this number.

Furthermore, we note that although the hottest stars in our NGC~104 sample (i.e. those with $T_{\rm eff}>$5800~K) have Li abundances in the range 1.64\,dex$<$ A(Li) $<$2.52\,dex, most of the stars cluster around A(Li)$\sim$1.9\,dex, which is significantly lower than most of the model predictions for A(Li)$_{i}$.
Even taking into account the correction of $-175$\,K applied to our stellar temperatures (in order to be consistent with the GES measurements), one would not reach the initial lithium we have inferred.
The lower Li abundances derived in these stars could be explained if these stars had already evolved past the TO. As we   noted  in Section 4, previous studies have classified these stars as TO \citep[][]{ dorazi10, dobrovolskas14}, but a closer look at our isochrone comparison in the $T_{\rm eff}$--M$_V$ plane (Figure \ref{fig:N2243-TeffvsMv}) indicates that they may already be in their SGB phase and may already have experienced some dredge-up and diluted their lithium content. However, even though this explanation is quite reasonable, we need to keep in mind that the exact evolutionary status of these stars remains uncertain.

In comparison, the NGC\,2243 stars in the same temperature range (5800\,K$ <T_{\rm eff}< $5900\,K) have Li abundances clustered around A(Li)$\sim$2.4\,dex, which is $\sim$0.5\,dex lower than what the model predicts for its initial Li ($\sim2.90\,$dex) and about 0.5\,dex higher than the stars in the same temperature range in NGC~104. This may hint at some Li depletion on the MS. When considering the large age difference between the two clusters, we can expect that the effect of diffusion is stronger in NGC\,104 than in NGC\,2243. On the other hand, the low Li abundance that we observe in the NGC\,104 stars (possibly subgiants) implies that the depleting mechanism cannot be simply diffusion if the initial Li of NGC\,104 was at the level predicted by the SBBN. Otherwise, these stars should have higher Li contents.

Finally, we note that the models closely reproduce the overall evolution of the Li content observed in both clusters, from the MS to the RGB bump. It is  now possible to  use our observationally derived abundances to provide important constraints to the model developers, by carrying out detailed comparisons of abundance levels and differences between key evolutionary stages. 
  However, our findings and feedback is currently based only on two clusters, both at rather high metallicities, whereas robust constraints will be tested on a broader parameter space (Aoki et al. 2021b, in prep.). 

\section{Conclusions}
We   compared the Li abundances in two large stellar samples of NGC~104 and NGC~2243. The two clusters have similar metallicities, but NGC\,104  is a globular cluster, while NGC\,2243 is an open cluster. 
More importantly, NGC2243 is one of the most metal-poor open clusters with main-sequence stars hotter than the Li dip, which can be used to derive its initial Li content \citep{randich2020gaia}. 
Our data samples include both UVES and FLAMES/GIRAFFE archival spectra, all publicly available. A subset of the NGC\,104 sample used in this work belongs to our own observing programme (Aoki et al. 2021b, in prep.).

We   followed the method introduced by \cite{mucciarelli12} and computed from the observed spectra the $\Delta$(Li) between the hot TO and pFDU stars in NGC\,2243. Subsequently, under the assumption of a similar evolutionary history in the two clusters under investigation because of their similar metallicity, we   applied this $\Delta$(Li) value to the Li observed in NGC\,104 pFDU stars to infer its initial Li abundance. 

For NGC\,2243, we  measured an initial lithium abundance of A(Li)$_i$=2.85 $\pm$ 0.09\,dex and derived $\Delta$(Li)=1.69 $\pm$ 0.13\,dex, which confirms, also based on comparisons with previous works, that $\Delta$(Li) increases with metallicity \citep[cf. ][]{mucciarelli12}. 
For NGC\,104, we   inferred an initial lithium abundance of A(Li)$_i$=2.34 $\pm 0.13$\,dex. This is significantly lower than the WMAP$+$BBN primordial Li value,  but in overall   agreement with the 
%lower end of the range of 
initial Li amounts found by \cite{mucciarelli12} in all globular clusters  observed to date, and in metal-poor field stars (Spite plateau). This result indicates that the Li abundance has slowly depleted during the MS phase or that the initial A(Li)$_i$ was lower than the predicted value of the BBN.

In order to further test our approach, and because we know that the difference in age between the two clusters implies different TO masses, hence possibly different amounts of depletion and different evolutionary histories, we  also compared our observed Li abundances to the  theoretical model predictions.

The models predict initial Li abundances of A(Li)$_{i}$=2.81$-$2.90\,dex (for NGC\,2243, depending on the model) and A(Li)$_i$=2.61\,dex (for NGC\,104), which show  very good agreement with the value empirically derived in the case of NGC\,2243 (A(Li)$_i$=2.85 $\pm$ 0.09\,dex), while in the case of NGC\,104 the two values are barely in agreement (A(Li)$_i$=2.61\,dex versus A(Li)$_i$=2.34 $\pm$ 0.13\,dex).
However, models computed for the same total metallicity (Z=0.008) and without diffusion predict a 0.22\,dex difference between the $\Delta$(Li) values of NGC\,2243 (1.74\,dex) and NGC\,104 (1.96\,dex).
We can then conclude that the model predictions validate our empirical approach, but can also offer further constraints on a possibly more realistic dilution factor for NGC\,104, which is based on our initial assumption that the two clusters have undergone a similar evolution of their lithium content because of their similar metallicity. However, because our models were not   computed specifically for this purpose or thoroughly tested, the 0.22\,dex difference between the theoretical dilution factors can only be considered a first-order correction. If applied to the pFDU A(Li) value, this  would imply an empirically derived initial lithium content in NGC\,104 of A(Li)$_i$=2.56 $\pm$ 0.13\,dex, in very good agreement with the model prediction.

Finally, in order to use our observational results to further improve the stellar models and constrain the different mixing mechanisms at work in stellar interiors and affecting lithium surface abundances, one would ideally need to properly sample the MS phase of NGC\,104. 

\begin{acknowledgements}
We acknowledge the use of the ESO Science Archive Facility and more specifically of the Gaia-ESO Public Spectroscopic Survey catalogue. This work has also made use of the BaSTI set of isochrones and models. We deeply thank N. Lagarde and C. Charbonnel for enriching  discussions and for sharing their expertise and stellar models, at an early stage of this work. We also thank the anonymous referee for their detailed comments that helped improving the paper.
MA acknowledges support from the IMPRS on Astrophysics at the LMU M{\"u}nchen. 

\end{acknowledgements}

% WARNING
%-------------------------------------------------------------------
% Please note that we have included the references to the file aa.dem in
% order to compile it, but we ask you to:
%
% - use BibTeX with the regular commands:
%   \bibliographystyle{aa} % style aa.bst
%   \bibliography{Yourfile} % your references Yourfile.bib
%
% - join the .bib files when you upload your source files
%-------------------------------------------------------------------
\bibliographystyle{aa}
\bibliography{47TucN2243}

\end{document}